\documentclass[aps,twocolumn,prc,superscriptaddress,showpacs,nofootinbib,floatfix,amssymb,amsfonts,amsmath]{revtex4-1}

\usepackage{graphicx}
\usepackage{dcolumn}
\usepackage{bm}
\usepackage{amssymb}
\usepackage{color}
\usepackage{float}
\usepackage{ltxcmds}

\begin{document}

\title{Anchor-based optimization of energy density functionals}

\author{A.\ Taninah}
\affiliation{Department of Physics and Astronomy, Mississippi
State University, MS 39762}

\author{A.\ V.\ Afanasjev}

\affiliation{Department of Physics and Astronomy, Mississippi
State University, MS 39762}

\date{\today}

\begin{abstract}
  A new anchor-based optimization method of defining the  energy density functionals (EDFs) 
is proposed.  In this approach, the optimization of the parameters of EDF is carried out for 
the selected set of spherical anchor nuclei the physical observables of which are modified by 
the correction function which takes into account the global performance of EDF.  It is shown  
that the use of this approach leads to a substantial improvement in global description of binding 
energies for several classes of covariant EDFs. The computational cost of defining a new 
functional within this approach is drastically lower as compared  with the one for the optimization 
which includes the global experimental data on spherical, transitional and deformed nuclei into 
the fitting protocol.
\end{abstract}

\maketitle

    Nuclear density functional theory (DFT) is presently one of the most widely 
used self-consistent approaches in low-energy nuclear physics 
\cite{BHP.03,VALR.05,Drut2010_PPNP64-120,PM.14}. It is based on the concept of 
energy density functional (EDF)  the several parameters of which are defined by the 
properties of finite nuclei and nuclear matter properties. This approach is universal in a
sense that it allows the global description of nuclear properties covering the nuclear 
landscape from light to very heavy nuclei and from known to unknown nuclei 
\cite{Eet.12,AARR.14}. It also provides important information, such as masses, decay
and fission rates etc, for nuclear astrophysics \cite{AG.20,CSLAWLMT.21}.

    However, the definition of EDFs is not unique and faces a number of challenges the
part of which is related  to fitting protocols \cite{DNR.14,AAT.19}. At present, absolute majority of 
the EDFs are fitted to the properties of spherical nuclei. This led to a huge number of the 
functionals the global performance of which is not established. There are more than 300 relativistic 
(covariant) EDFs (further CEDFs) (see, for example, Ref.\ \cite{RMF-nm}) and comparable number 
of non-relativistic Skyrme EDFs.  However, such an approach creates a substantial bias towards 
spherical nuclei: the improvement of the functional for spherical nuclei frequently leads to 
a degradation of its global performance. Only very limited number of non-relativistic Skyrme and 
Gogny EDFs and  only one CEDF have been fitted globally to experimental  data which includes 
spherical, transitional and deformed nuclei (see Refs.\ 
\cite{GCP.09,GCP.13,UNEDF0,UNEDF1,GHGP.09,AGC.16}). However, the computational
cost of the generation of such functionals is enormous. 

    In the present paper a new method of anchor-based optimization of the functionals is 
proposed in order to alleviate these problems.  It  combines the  simplicity of the optimization 
of EDFs to spherical nuclei with the information on their  global  performance. In contrast to 
global fits of EDFs, it typically requires only several rounds of global calculations to
achieve a significant improvement of global performance of CEDFs. This  was verified for 
several classes of CEDFs and achieved at a moderate increase of computational time as
compared with the optimization to only spherical nuclei.


    The general procedure for the anchor-based  optimization of EDFs is the following:

\begin{enumerate}

\item
The set of "anchor" spherical nuclei is selected and the optimization of the functional
is carried out with spherical relativistic Hartree-Bogoliubov (RHB)  computer code 
using experimental data on these nuclei and nuclear matter properties  (NMP) (see 
Refs.\ \cite{AARR.14,TAAR.20} and Supplementary Material  \cite{Suppl-anchor} for details). The 
obtained functional is labelled as EDF$_{i}$ ($i=0$). Here $i$ is the counter 
of  the iteration in the anchor-based optimization.

\item 
 Global calculations of the masses, charge radii and other physical observables are carried out with 
axially deformed RHB code using EDF$_{i}$ for the set  of nuclei  in which respective experimental 
data exist.  The set of binding energies $E_{EDF_i}(Z,N)$ is defined for the $n$ even-even nuclei the 
masses of which have been either measured or estimated in the AME2016 mass evaluation 
\cite{AME2016-third}. Note that this set of the nuclei includes spherical, transitional and deformed 
nuclei.

\item
The correction function  
\begin{eqnarray}  
E_{corr}(Z,N) = \alpha_i(N-Z) + \beta_i(N+Z) + \gamma_i
\label{Eq-1} 
\end{eqnarray} 
is added\footnote{Alternative functional dependencies have been explored. However,
Eq.\ (\ref{Eq-1}) brings the best improvement of the EDFs.}
 to the obtained set of calculated binding energies
\begin{eqnarray}
E_{pseudo}(Z,N) = E_{EDF_i}(Z,N) + E_{corr}(Z,N)
\end{eqnarray}
Then, the optimal parameters $\alpha_i$, $\beta_i$ and $\gamma_i$ are determined by 
minimizing $\Delta E_{rms}$ defined as:
\begin{eqnarray}
\Delta E_{rms} = \sqrt{\frac{\sum_{k=1}^{n} (E_{pseudo}(Z,N) - E_{exp}(Z,N))^2}{n}} 
\end{eqnarray} 
where $E_{exp}(Z,N)$ is experimental binding energy of the $k$th nucleus with $(Z,N)$
and $k$ runs over all even-even nuclei for which experimental data exist.
Thus, the addition of $E_{corr}(Z,N)$ to binding energies aims at the minimization
of global difference between calculated and experimental binding energies.
  The variation of the parameters $\alpha_i$, $\beta_i$ and $\gamma_i$ 
with iteration number during the iterative procedure are illustrated in
Tables 1-3 of Supplemental Material \cite{Suppl-anchor}.

\item
The energies of spherical anchor nuclei are redefined as
\begin{eqnarray}
E^{pseudo}_{exp}(Z,N) = E_{exp}(Z,N) + E_{corr}(Z,N)
\end{eqnarray}
where $E_{corr}(Z,N)$ is calculated for optimal  parameters $\alpha_i$, $\beta_i$ and $\gamma_i$
defined in previous step. New set of parameters EDF$_{i+1}$ is defined using 
$E^{pseudo}_{exp}(Z,N)$ as "experimental" data and the procedure of the point (1).

\item
 New global calculations are carried out using  EDF$_{i+1}$ and the procedure of point (2). 
A significant improvement in the global description of masses has been achieved at
 the first  step of iterative procedure for the DD-MEY, NL5(Y) and PC-Y functionals and
 at  third step for the DD-MEX1 one\footnote{Such feature of the anchor  based optimization
 method can be extremely useful for theoretical groups with limited computational
 resources.}. Further repetition of the steps (3)-(5) leads to only moderate improvement
 of the global description of the masses. 
 
   The convergence of iterative procedure is reached in the limit $\alpha_i \rightarrow 0$, 
 $\beta_i \rightarrow 0$, and $\beta_i \rightarrow 0$. However, the analysis presented
 in Tables 1-3 of Supplemental Material \cite{Suppl-anchor} clearly suggests that the
 termination of iterative procedure before reaching this limit may lead to only very small
 (if any) degradation of the global description of masses.

 
  
\end{enumerate}

\begin{table*}[htb]
  \caption{
    The rms deviations $\Delta E_{rms}$,
    $\Delta (S_{2n})_{rms}$,
    $\Delta (S_{2p})_{rms}$, and $\Delta (r_{ch})_{rms}$
between calculated and experimental binding energies $E$, two neutron (two-proton) separation energies $S_{2n}$ ($S_{2p}$), and 
charge radii $r_{ch}$. The first three observables  are determined with respect of  "measured + estimated" set of experimental masses 
of 855 even-even nuclei  from the AME2016 mass evaluation \cite{AME2016-third}. The $\Delta (r_{ch})_{rms}$
values are calculated using  experimental data on charge radii of 305 even-even nuclei from Ref.\ \cite{AM.13}.
The values shown in parenthesis ( ) are the rms deviations for the subset of nuclei which excludes light nuclei
with $A<70$. The incompressibility $K_{0}$, the symmetry energy  $J$, and the slope of the symmetry energy $L_{0}$ 
of the functionals under study are shown in columns 6, 7 and 8, respectively.
\label{table-res}
}
\begin{tabular}{|  c| c| c| c| c| c| c| c|}
  \hline  
                             &  $\Delta E_{rms}$   &$\Delta (S_{2n})_{rms}$ & $\Delta (S_{2p})_{rms}$ & $\Delta (r_{ch})_{rms}$  & $K_{0}$ & $J$ &$L_{0}$ \\
                             &  [MeV]    & [MeV]         & [MeV]        & [fm]          &    [MeV]    & [MeV]     &   [MeV]      \\
                              &              &                   &                  &                &                &               &                \\\hline              
 1 & 2 & 3 & 4 & 5 & 6 & 7&8     \\ \hline                                                                                                                                                                                   
 \multicolumn{8}{|c|}{} \\ \hline                                                                                                                                                                                     
 DD-ME2 \cite{DD-ME2}        &    2.436 (2.300)      & 1.056 (0.854)         &    0.949 (0.750)       & 0.0266 (0.0262)  & 250.9 & 32.9 & 49.4  \\
 DD-MEX \cite{TAAR.20}       &    2.849 (2.963)      & 1.095 (0.972)         &    0.978 (0.847)       & 0.0247 (0.0249)  & 267.0 & 32.9 & 47.8   \\
 DD-MEX1                     &    1.637 (1.539)      & 1.045 (0.873)         &    0.896 (0.704)       & 0.0261 (0.0263)  & 291.8 & 32.5 & 51.8  \\ \hline
 \multicolumn{8}{|c|}{} \\ \hline                                                                                                                                                                                            
 DD-MEX2                     &    2.236 (1.791)      & 1.228 (0.913)         &    1.271 (0.928)       & 0.0466 (0.0488)  & 255.8 & 35.9 & 85.3          \\
 DD-MEY                     &    1.734 (1.414)      & 1.259 (0.876)         &    1.026 (0.755)       & 0.0264 (0.0244)  & 265.8 & 32.8 & 51.8          \\ \hline
 \multicolumn{8}{|c|}{} \\ \hline                                                                             
 NL5(E) \cite{AAT.19}          &    2.802 (2.689)      & 1.204 (0.864)         &    1.366 (1.033)       & 0.0285 (0.0271)  & 253.0 & 38.9 &125.0           \\
 NL5(Y)                       &    2.362 (1.675)      & 1.256 (0.709)         &    1.222 (0.772)       & 0.0297 (0.0292)  & 254.5 & 36.6 &116.7          \\ \hline
  \multicolumn{8}{|c|}{} \\ \hline                                                                            
 PC-PK1 \cite{PC-PK1}        &    2.400 (2.149)      & 1.331 (0.932)         &    1.354 (0.875)       & 0.0306 (0.0269)  &  238     &  35.6    &  113    \\
 PC-Y                       &    1.951 (1.600)      & 1.438 (0.770)         &    1.175 (0.690)       & 0.0311 (0.0247)  &  241      & 35.1     & 105          \\
 PC-Y1                     &    1.849 (1.509)      & 1.345 (0.846)         &    1.106 (0.822)       & 0.0294 (0.0249)  &  240      & 34.9     &  107       \\ \hline 
\end{tabular}
\end{table*}

   The spherical and deformed calculations are carried out using the RHB computer
codes developed in Refs.\ \cite{TAAR.20,AARR.14}.  The truncation of the basis is performed in 
such a way that all states belonging to the major shells up to $N_F = 20$ fermionic shells for the 
Dirac spinors and up to $N_B = 20$ bosonic shells for the meson fields are taken into account. 
Note that the latter applies only to the functionals which contain meson exchange such as those
belonging to non-linear meson couplings (NLME) and density dependent meson-nucleon coupling
(DDME)  classes of the functionals (see Refs.\ \cite{AARR.14,TAAR.20}). The accuracy of the
truncation of the basis is discussed in Supplemental Material \cite{Suppl-anchor}.

   In order to avoid the uncertainties connected with the definition of the size of the pairing window 
the separable form of the finite range Gogny pairing interaction introduced by Tian et al \cite{TMR.09} 
is used with two versions of the strength $f$ of the pairing. In the first one (called further as "Pair-1"), 
the pairing strength is dependent on proton number (see Ref.\ \cite{AARR.14} for detail). In the second 
one called "Pair-2" (see Ref.\ \cite{TA.21} for detail), the proton pairing is made mass dependent via 
\begin{eqnarray}
f_\pi = 1.877 (N+Z)^{-0.1072},
\label{scaling-factors-pr}
\end{eqnarray}
and neutron pairing is isospin dependent via 
\begin{eqnarray}
f_\nu = 1.208 e^{-0.674\frac{|N-Z|}{N+Z}}.
\label{scaling-factors-nu}
\end{eqnarray} 
This type of phenomenological scaling of pairing strength provides the best reproduction 
of the experimental pairing indicators (see Ref.\ \cite{TA.21}). Note that the labels of the 
functionals defined with "Pair-2" pairing contain the letter "Y" at/near their end.

     The anchor-based optimization method is applied here for the 
DDME, NLME  and  point coupling (PC) classes of CEDFs (see Refs.\ \cite{AAT.19,TAAR.20} 
and Supplemental Material \cite{Suppl-anchor}  for technical details). 
For each class, the functionals with best global performance
such as DD-ME2 \cite{DD-ME2}, NL5(E) \cite{AAT.19} and
PC-PK1 \cite{PC-PK1} are used as a starting point.  
In addition, the basic features 
of their fitting protocols are employed here.  As a consequence, there are 12 anchor 
spherical nuclei in the DDME and NLME models and 60  anchor spherical nuclei in the 
PC model. The types of the input data for the fitting protocols and related adopted errors 
are summarized in Table 4 of the Supplemental Material \cite{Suppl-anchor}.

     The optimization of the CEDFs in spherical nuclei is performed in the following
way.  First, approximately 200 trials of minimization from  the sets of initial 
parameters, randomly generated in large parameters hyperspace, are performed 
using the simplex based minimization method.  Second, the minimization is 
repeated by using initial parameters generated in smaller parameter hyperspaces
around several local minima characterized by the lowest penalty functions
using both simplex-based and simulated annealing\footnote{It is our experience
that simulated annealing method is extremely costly and numerically unstable 
for large parameter hyperspaces.}  minimization methods. This procedure
guarantees the convergence to the global minimum and provides 
information on parametric correlations between the parameters of
CEDFs (see Ref.\ \cite{TAAR.20}).
  
    The global performance of existing (DD-ME2, NL5(E) and PC-PK1)  and 
new (DD-MEX1, DD-MEX2, DD-MEY, NL5(Y), PC-Y and PC-Y1) functionals obtained by means 
of anchor-based optimization method are summarized in Table \ref{table-res}. When 
considering the quality of the functionals one should take  into account the ranges of the 
nuclear matter properties (NMPs) recommended for relativistic functionals in Ref.\ \cite{RMF-nm}. These are 
$\rho_0 \approx 0.15$ fm$^{-3}$, $E/A \approx -16$ MeV, $K_0= 190-270$, $J=25-35$ 
MeV ($J=30-35$ MeV) and $L_0=25-115$ ($L_0=30-80$) for the SET2a (SET2b) sets of 
the constraints on the experimental/empirical ranges for the quantities of interest.
  
    The DD-MEX1 functional originates from the DD-ME2 one: both have the 
same fitting protocol (see Supplemental Material \cite{Suppl-anchor}) but the CEDF DD-ME2 is fitted at the 
BCS level employing monopole pairing while the DD-MEX1 one at the RHB level with 
the "Pair-1" separable pairing.  However, the DD-MEX1 functional, fitted with anchor-based
optimization method,  provides a substantial improvement in the global description of binding
energies (from $\Delta E_{rms} = 2.436$ MeV for DD-ME2 down to $\Delta E_{rms} = 1.651$ 
MeV for DD-MEX1, see Table  \ref{table-res}). It also provides a slight improvement in the 
description of two-neutron and two-proton separation energies and charge radii. The most of 
the NMPs of this functional are within the SET2b limits: the only exception is incompressibility 
$K_0$ which exceeds the  SET2b upper limit.

   It is interesting to see whether the binding energies and charge radii alone can 
provide a reasonable constraint on NMPs and neutron skins. For that, the DD-MEX2
and DD-MEY functionals have been created the fitting protocols of which do not
contain any information on NMPs and neutron skins (see Table 4 in Supplemental
Material \cite{Suppl-anchor}). In addition, the adopted errors for binding energies are 
fixed at 1.0 MeV for all nuclei in this class of the functionals.  The DD-MEX2 functional 
has been optimized with the "Pair-1" separable pairing. As compared with DD-ME2 
(DD-MEX1)   it leads to some improvement (degradation) in the description of binding energies 
but  provides  less accurate description of two-proton and two-neutron separation
energies and charge radii as compared with two above mentioned functionals
(see Table  \ref{table-res}).

    The situation drastically improves when the "Pair-2" separable pairing is used. This  leads 
to the DD-MEY functional which provides second best global description of binding 
energies  ($\Delta E_{rms} = 1.734$ MeV) and the best description of the binding energies
of the $A>70$ nuclei ($\Delta E_{rms} = 1.414$ MeV) among considered functionals [see Table  
\ref{table-res}] 
 Despite the fact that the DD-MEY functional was fitted without constraint on 
NMPs,  they are within the SET2b limits (see Table  \ref{table-res}). In particular, it gives a more
reasonable value of incompressibility $K_0$ than the DD-MEX1 functional. 
These facts potentially indicate the importance of the isospin dependent neutron pairing in the 
simultaneous description 
of binding energies and NMP.
However, the 
description of two-neutron and two-proton separation energies with this functional
is somewhat worse as compared with the DD-MEX1 one.

   To verify the results obtained with anchor based optimization method we also 
employed the method of minimization of Ref.\ \cite{TGPO.00} which is used
in nuclear mass table fits by Brussels group. In this method,  the binding energies 
of deformed and transitional nuclei are corrected  by the deformation energies so 
the optimization is carried out for the energies of spherical solution of the nuclei used in the
fitting protocol. In a given nucleus, the 
deformation energy represents the difference between the  energies of the global 
minimum with deformations $\beta_i \neq 0\qquad (i=2, 4,...)$ and spherical 
solution with  $\beta_i=0$. Because of available computers we used 400 even-even 
nuclei evenly spread over nuclear chart [starting from actinides and going down to 
light nuclei and eliminating each second even-even nucleus] in these calculations.
The iterative procedure in this method requires new calculations of deformation
energies at each iteration (see Ref.\ \cite{TGPO.00} for detail) and it turns out that 
their convergence is quite slow especially in the "DD-MEX1" type of the functional.  
The rms deviations between experimental and calculated binding energies 
$\Delta E_{rms}$ obtained in these calculations are 1.672(0.068)\footnote{The numbers 
shown in parentheses  are the errors in deformation energies due to limited number of 
iterations in the iterative procedure.} and 
1.613(0.258) MeV for the "DD-MEY" and "DD-MEX1" functionals, respectively. 
Thus, the results obtained with this approach are in line with those obtained in
anchor based optimization approach (see Table \ref{table-res}); some difference
in the results are due to different selection of the nuclei directly included into
fitting protocol.  Note that this approach is numerically substantially more time 
consuming as compared with anchor based optimization approach because
(i) substantially more "spherical" nuclei (400 versus 12) are used in minimization
procedure and (ii) slow convergence of deformation energies in the iterative
procedure.

   The anchor-based optimization method has been applied also to the NLME and 
PC classes of the CEDFs. In both cases, it leads to an improvement of global description of masses. 
The NL5(E) functional (see Ref.\ \cite{AAT.19}) 
is the starting point for the optimization of the NL5(Y) CEDF.  Note that the "Pair-1" and "Pair-2" 
separable pairing is used in the fitting protocols of the  NL5(E) and NL5(Y) CEDFs, respectively.
Table \ref{table-res} shows that the application of anchor based optimization method in
combination with the use of isospin dependent neutron pairing leads to a substantial 
improvement of global mass description and some improvement of the symmetry energy
$J$ and the slope of the symmetry energy $L_0$. The same situation exists also for the
PC functionals.  The anchor based optimization starting from the PC-PK1 functional leads
to the CEDF PC-Y which provides a substantial improvement of the global description of 
binding energies (see Table \ref{table-res}).

\begin{figure*}[htb]
\centering
\includegraphics*[width=8.5cm]{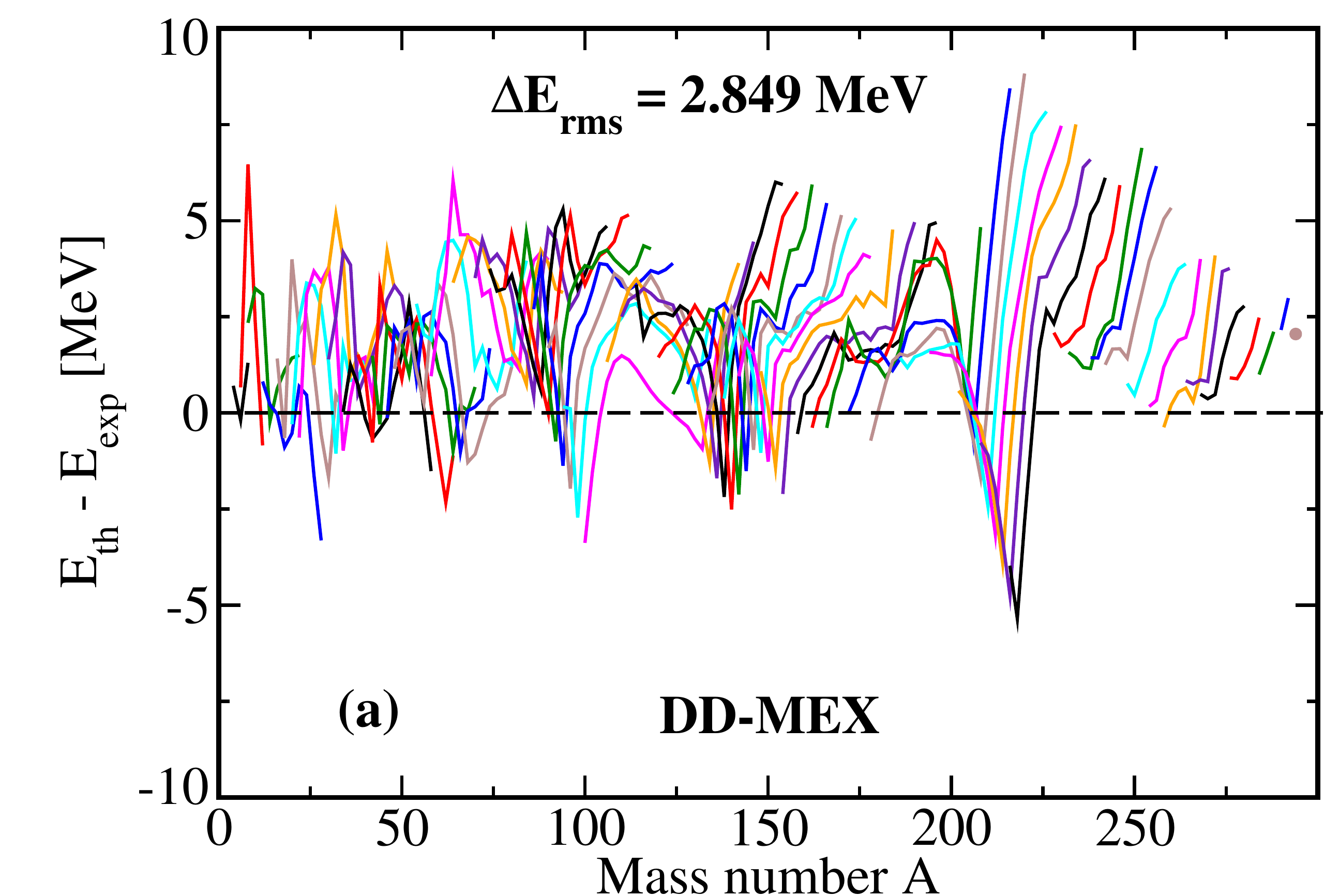}
\includegraphics*[width=7.5cm]{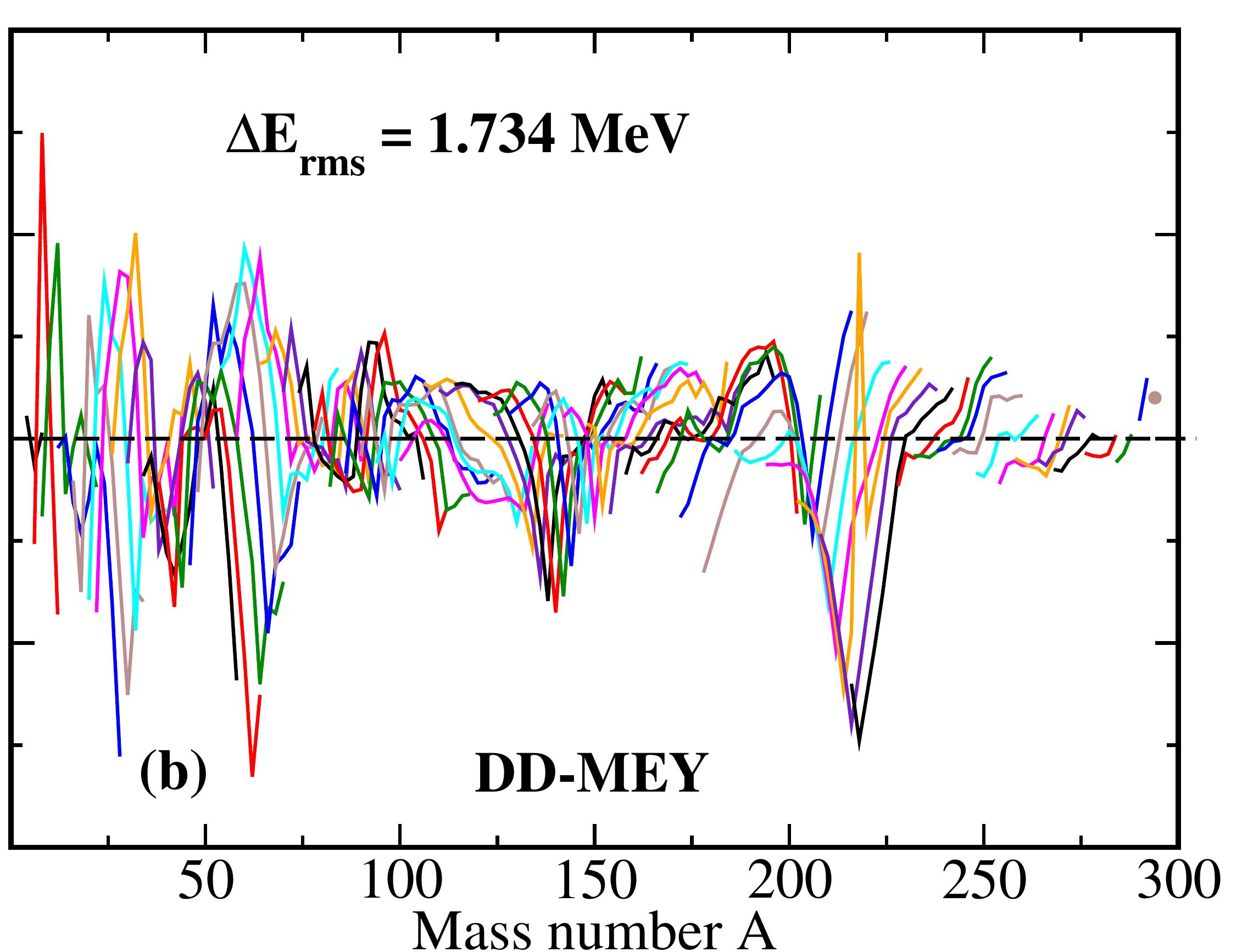} \\
\hspace{0.75cm}
\includegraphics*[angle=-90,width=8.2cm]{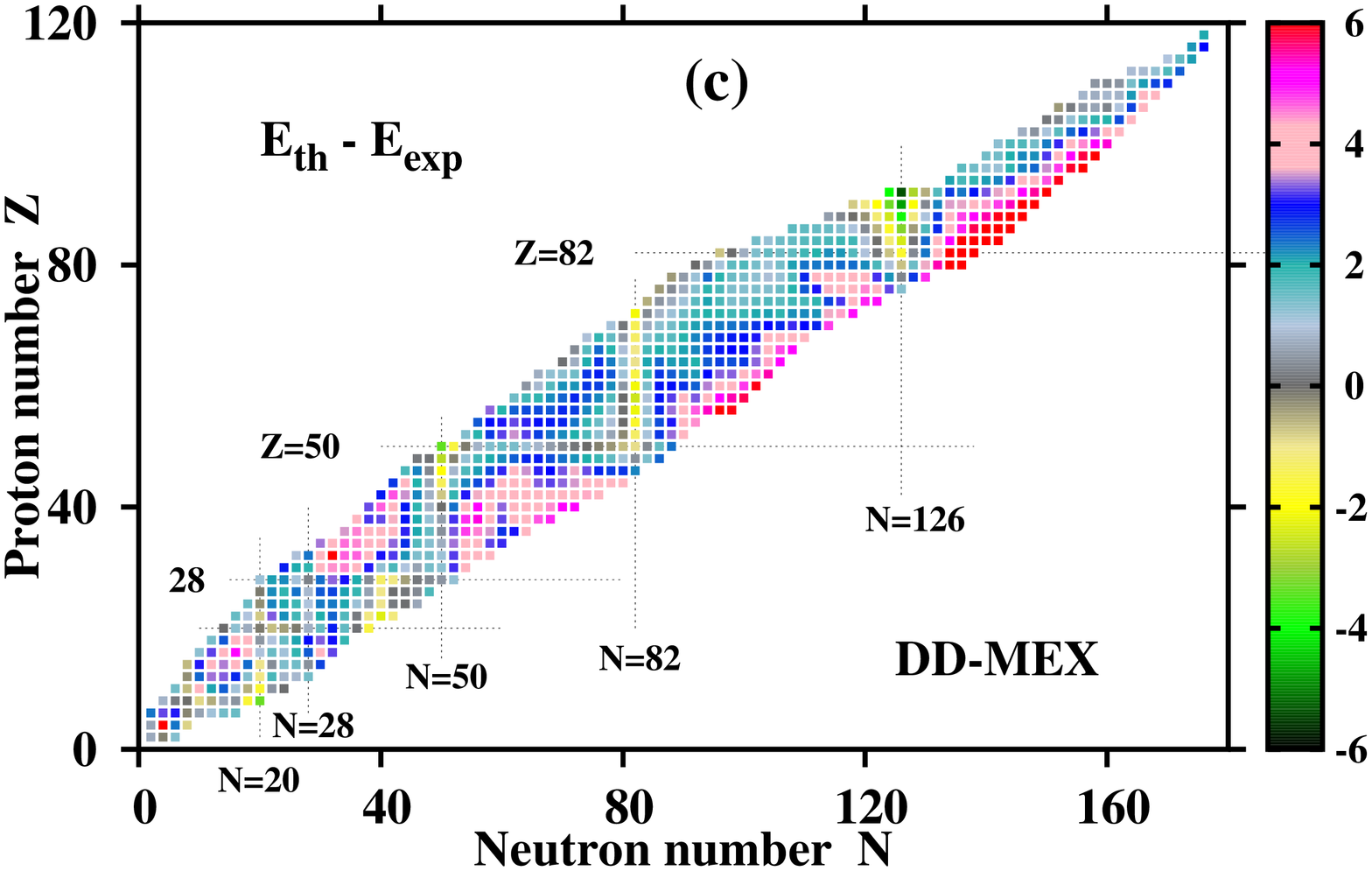}
\includegraphics*[angle=-90,width=8.2cm]{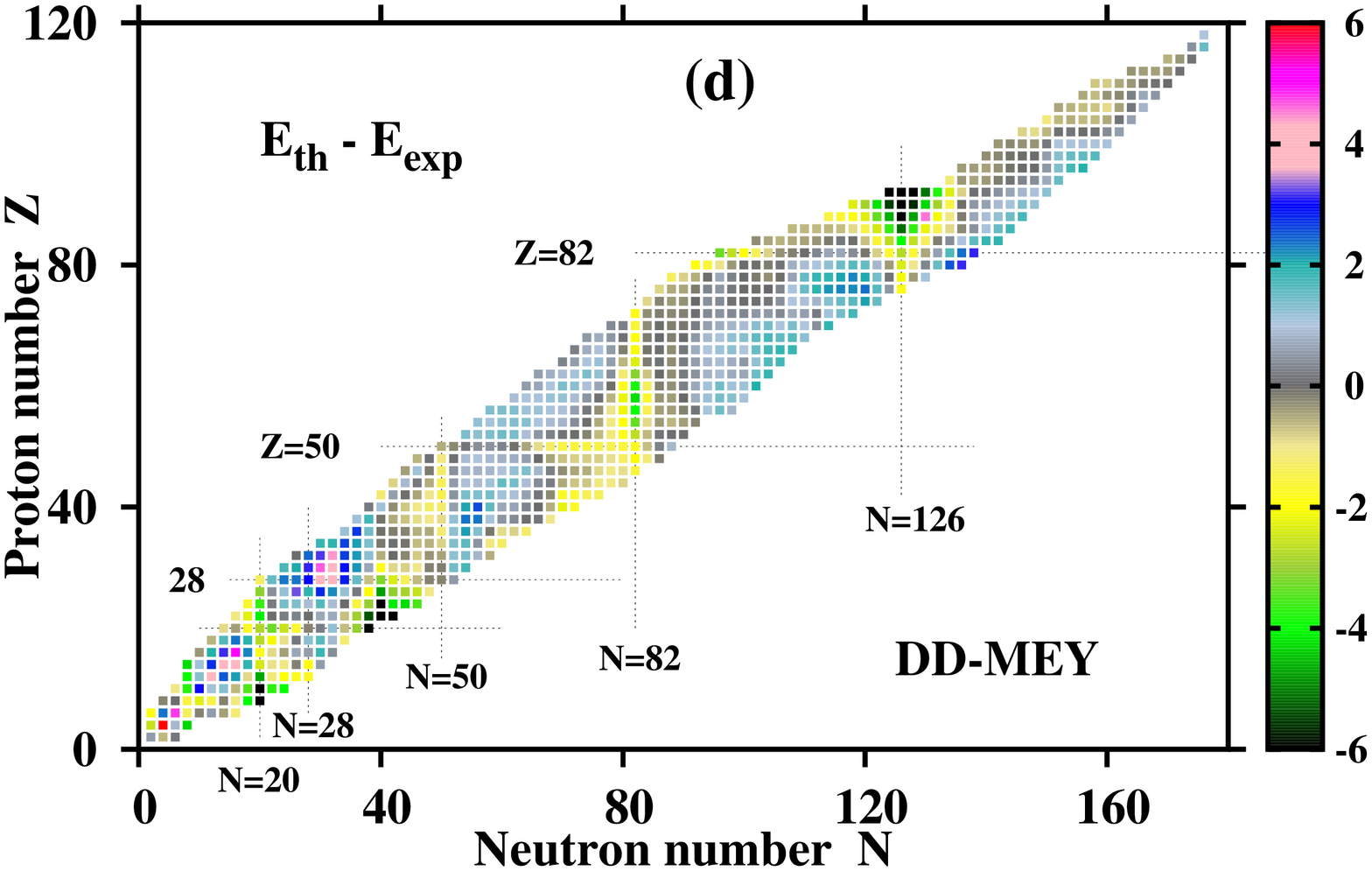}
\caption{The differences $E_{th} - E_{exp}$ between calculated ($E_{th}$)  
and experimental ($E_{exp}$) masses for indicated CEDFs. All 855 even-even 
nuclei, for which measured and estimated masses are available in the AME2016 
compilation \cite{AME2016-third}, are used in this comparison.
If $E_{th} - E_{exp} < 0$, the nucleus is more bound in the 
calculations than in experiment.
\label{fig-deviations}
}
\end{figure*}

    Note that the fitting protocols of above mentioned DDME and NLME 
functionals include only 12 spherical anchor nuclei. For these functionals it was verified 
that the increase of the number of spherical anchor nuclei to 60 (as in the fitting protocol 
of the PC-PK1 functional \cite{PC-PK1}) does not lead to an improvement of the global 
description of binding energies.  The same conclusion is valid for the PC model: the
reduction of the number of the anchor spherical nuclei from 60 in the PC-Y functional down to 12 in 
the  PC-Y1 functional (see Table 1 in Supplemental Material \cite{Suppl-anchor}) leads to some
improvement in the description of physical observables (see Table \ref{table-res}). 

    With the exception of the DD-MEX2 functional all CEDFs shown in Table \ref{table-res} 
give comparable rms deviations for charge radii $\Delta (r_{ch})_{rms} \approx 0.026$ fm, corresponding 
to a high precision of $\approx 0.5\%$ in charge radii predictions.  These results and the analysis
of Ref.\ \cite{PAR.21} suggest that the inclusion of global data on charge radii will not likely lead to
an appreciable improvement of the functionals.

     Fig. \ref{fig-deviations} illustrates the improvements in the global description of the 
masses and related physical observables when the anchor-based optimization method is 
employed. DD-MEX is the best DDME  functional as defined by the penalty function 
of the fitting protocol including only spherical nuclei (see Ref.\  \cite{TAAR.20}). 
 However, this bias towards spherical nuclei leads to  $\Delta E_{rms} = 2.849$ 
MeV in global description of the masses (see Table \ref{table-res}) and appreciable deviations 
between theory and experiment displayed in Figs.\ \ref{fig-deviations}(a) and (c).  In particular, it 
leads to a systematic shift of the average $E_{th}-E_{exp}$ values from the $E_{th}-E_{exp}=0$ 
line [see Fig.\ \ref{fig-deviations}(a)]. In contrast,  such a shift does not appear for the DD-MEY
functional [see Fig.\ \ref{fig-deviations}(a)]  which, in addition, improves the description of isospin 
dependence of the $E_{th}-E_{exp}$  values [compare Figs.\ \ref{fig-deviations} (c) and (d)].

    Note that these improvements are obtained  by a larger emphasis on medium and 
heavy mass  nuclei with $A>70$ in the anchor-based optimization method. For these nuclei
a substantial improvement in the description of the masses and two-particle separation energies
is obtained [see Table \ref{table-res} and compare Figs.\ \ref{fig-deviations} (a) and (b) and 
Figs.\ \ref{fig-deviations} (c) and (d)]. In contrast, the spreads in the  $E_{th}-E_{exp}$
values are getting larger for the light ($A<70$) nuclei [compare Figs.\ \ref{fig-deviations}(a) 
and (b)]. However, this is not critical since beyond mean field effects are expected to be 
larger in light ($A<70$) nuclei as compared with heavier ones. This is in line with the 
observation that the accuracy of the description of the masses and two-particle separation 
energies improves substantially when light $A<70$ nuclei are excluded from the analysis (see 
Table \ref{table-res}).  This improvement is especially drastic for the DD-MEX1, DD-MEX2, 
DD-MEY, NL5(Y), PC-Y and PC-Y1 functionals defined by the anchor-based optimization 
method.

   It is interesting to compare the performance of these functionals with those obtained
in Skyrme DFT for the UNEDF* class of the functionals which similar to our approach
has been defined at the mean field level without inclusion of rotational and vibrational
correlations. The UNEDF0 \cite{UNEDF0}, UNEDF1 \cite{UNEDF1}, and UNEDF2 
\cite{UNEDF2} EDFs were optimized by fitting their parameters to large (but restrictive) 
sets of experimental data involving spherical and deformed nuclei. These EDFs 
describe globally nuclear masses with $\Delta E_{rms} = 1.428$ MeV (UNEDF0), 
1.912 MeV (UNEDF1), and 1.950 MeV (UNEDF2). These values are close to those
obtained with the DD-MEX1, DD-MEY, PC-Y and PC-Y1 CEDFs but numerical cost of
the optimization of these Skyrme EDFs is drastically larger than that in the anchor-based 
optimization method.

      In conclusion, a new anchor-based optimization method of defining the  energy 
density functionals has been proposed. It combines the simplicity of the fitting of 
EDF to spherical nuclei with global information on the reproduction of experimental
masses by EDF.  This is done by correcting the binding energies of the anchor
spherical nuclei used in optimization.  As a consequence, the computational cost
of defining a new functional is drastically lower as compared with alternative methods 
of the optimization which simultaneously includes the experimental data on  spherical 
and deformed nuclei. 
Despite that the global performance of the functionals,
defined by the anchor-based optimization method, becomes comparable with the one 
obtained for the UNEDF* class of non-relativistic Skyrme functionals.  
Although anchor-based method is applied here for CEDFs, it can also be used
for non-relativistic Skyrme and Gogny functionals.

     The functionals studied in the present paper are restricted to the ones defined at the 
mean field level. However, the anchor-based optimization method can be easily generalized 
to the approaches which include correlations beyond mean field.   For that the RHB approach 
in the point (2) of the anchor-based optimization method procedure has to  be replaced by a 
respective beyond mean field method (such as five-dimensional  collective Hamiltonian 
\cite{NLVPMR.09,LNVMLR.09,SALM.19})\footnote{Alternatively, one can use a more simplistic 
approach and add phenomenological rotational corrections to the binding energies calculated
in the RHB or similar non-relativistic approach: this is done in a number of the calculations of masses 
(see, for example, Refs.\ \cite{MNMS.95,HFB14-BSk14,ZNLYM.14}).}. This will allow to bypass 
the existing challenge of extreme computational cost  of fitting EDF at the beyond mean field 
level and generate such functionals.  It is reasonable to expect that they will lead to a 
further improvement of the description of binding energies (see, for example, Refs.\ 
\cite{ZNLYM.14,YWZL.21}).

   Note that in addition to the references directly cited in the main body of the manuscript
the Supplemental Material \cite{Suppl-anchor} provides citations to Refs. 
\cite{Niksic2014_CPC185-1808,AKR.96,AATG.19,NVR.08,NL3*,PREX-II.21,48Ca-neutron-skin.22,AMKBPCD.20,48Ca-208Pb-neutron-skin-abinitio.22}.

\bibliography{references-40-anchor-based-rev.bib}

\end{document}